# Toward Ka Band Acoustics: Lithium Niobate Asymmetrical Mode Piezoelectric MEMS Resonators

Yansong Yang, Ruochen Lu, Tomas Manzaneque, and Songbin Gong
Department of Electrical and Computer Engineering
University of Illinois at Urbana-Champaign
Urbana-Champaign, USA
yyang165@illinois.edu

*Abstract*— This work presents a new class of micro-electro-mechanical system (MEMS) resonators toward Ka band (26.5-40 GHz) for fifth-generation (5G) wireless communication. Resonant frequencies of 21.4 and 29.9 GHz have been achieved using the fifth and seventh order asymmetric (A5 and A7) Lamb-wave modes in a suspended Z-cut lithium niobate (LiNbO$_3$) thin film. The fabricated device has demonstrated an electromechanical coupling ($k_t^2$) of 1.5% and 0.94% and extracted mechanical $Q$s of 406 and 474 for A5 and A7 respectively. The quality factors are the highest reported for piezoelectric MEMS resonators operating at this frequency range. The demonstrated performance has shown the strong potential of LiNbO$_3$ asymmetric mode devices to meet the front-end filtering requirements of 5G.

## I. INTRODUCTION

Recently, due to the market demand for higher data rate for cellular phone applications, below 3 GHz spectrum has become increasingly crowded with little available spectrum for the expansion of fifth-generation (5G) wireless communication systems. As a result, Federal Communications Commission (FCC) has opened frequency bands 27.5-28.35 GHz and 37-40 GHz for licensed use, which overlaps with Ka band, for 5G mobile radio applications [1]. Several European and Asian countries have also planned portions of Ka band for the same purpose [2]. This expansion has sparked new technology development to overcome the remaining bottlenecks in accessing this portion of the spectrum. One key missing piece in beyond-24 GHz front-ends is high-performance and miniature filters that can be arrayed in a smart phone. At fourth-generation (4G) frequency ranges (<3 GHz), filters based on acoustic and MEMS resonators have been widely adopted. However, it remains challenging to scale these devices in frequency to meet 5G requirements.

The state-of-the-art acoustic resonators beyond 24 GHz resort to either the fundamental FBAR mode in a 100 nm thin AlN film or an overtone FBAR mode in a thicker AlN film [3] [4]. In either case, significant performance compromise has to be made in the process of scaling, which might lead to inadequate performance for 5G applications. For instance, a suspended 100 nm AlN thin film would produce a high thermal resistance to the surrounding. In addition, matching to system resistance (50 Ω) beyond 24 GHz requires small static capacitance ($C_0$<150 fF), which limit the size (length·width < 200 μm$^2$) of 100 nm thick

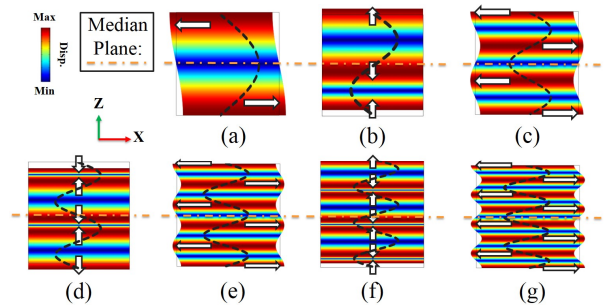

*Figure 1:* Displacement mode shapes of the (a) first-order (A1), (b) second-order (A2), (c) third-order (A3), (d) fourth-order (A4), (e) fifth-order (A5), (f) sixth-order (A6), and (g) seventh-order (A7) asymmetric modes. The arrows denote the displacement directions. The stress field for each mode order is plotted in black dashed line.

FBAR. All these limitations lower the power handling and increase the thermal nonlinearity of the resonator [5]. On the other hand, overtone operations, despite providing better power handling via a thicker film, would lead to diminished electromechanical coupling ($k_t^2$) as it will be discussed in this paper.

To overcome these limitations, one alternative is to select an acoustic wave mode in a material that features much higher electromechanical coupling than AlN FBARs so that frequency scaling toward Ka Band based on higher orders can still produce reasonable $k_t^2$ to meet 5G bandwidth requirement. It is also preferable that the mode has a large acoustic velocity so that scaling would not reduce the feature size to the extent that limits power handling. Recently, several modes (S0, SH0, and A1) have been demonstrated in LiNbO$_3$ with high coupling factors [6-10]. Among these wave modes, A1 Lamb wave mode has the highest phase velocity, and has been demonstrated with very large $k_t^2$ and $Q$s simultaneously in Z-cut LiNbO$_3$ ($k_t^2$=28% and $Q$=500 at 5 GHz), and in Y-cut LiNbO$_3$ ($k_t^2$=6.3% and $Q$=5341 at 1.7 GHz) [9] [10]. Therefore, A1 mode devices are being considered as an alternative resonator technology for sub-6 GHz applications. However, the higher order asymmetrical modes have not yet been explored for scaling the resonant frequency beyond 6 GHz.

In this work, we aim to extend asymmetrical Lamb wave modes in Z-cut LiNbO$_3$ to higher orders for beyond-24 GHz applications. We first investigate the excitation of asymmetric modes of various orders. More specifically, the dependence of electromechanical coupling factor on mode order and resonant frequency is studied with both analytical and finite element methods. Based on these studies, a film thickness of 400 nm is chosen to support the fifth and seventh order asymmetric (A5 and A7) modes to move up and reach Ka band. The fabricated device is measured with $k_t^2$ of 1.5% and 0.94% and extracted mechanical $Q$s of 406 and 474, at 21.4 GHz and 29.9 GHz for A5 and A7 modes, respectively.

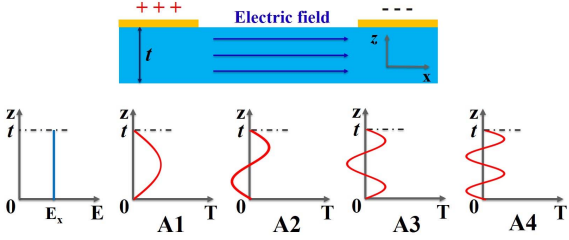

*Figure 2: Distribution of electrical and stress fields of A1, A2, A3, and A4 modes.*

## II. ASYMMETRIC LAMB WAVE MODES IN LITHIUM NIOBATE THIN FILM

Asymmetric modes are a class of Lamb-wave modes characterized by their particular anti-symmetry about the median plane of the plate. In other words, they have equal vertical displacement components but opposite longitudinal components on different sides of the median plane [11]. To visualize the displacement mode shapes of asymmetric modes of various orders, Comsol finite element analysis (FEA) is used to simulate the eigen modes in a 2D LiNbO$_3$ slab with free top and bottom surfaces and periodic boundaries in the lateral direction (Fig. 1). Various order modes from the first (A1) to seventh (A7) are shown with a mode order denoting the number of half-wavelength periodicities in the vertical direction. Clearly, overmoding a slab with a fixed thickness would yield a higher resonant frequency, provided the intended higher order mode can somehow be excited in the LiNbO$_3$ slab with transducers.

The simplest transducers for such a purpose are interdigital electrodes that are patterned exclusively on top of a transferred LiNbO$_3$ thin film, as they have least fabrication complication. As demonstrated by several prior works on A-modes, top-only interdigital transducers (IDT) can effectively excite the A1 modes with lateral electric fields [9] [10]. However, electromechanical coupling to asymmetric modes with mode orders higher than A1 has not been well studied with top-only IDTs. To understand the excitation of higher-order A-modes with IDT, magnitudes of the stress standing waves for different order modes are depicted in Fig. 1 and 2. As seen in Fig. 2, we simplify the E-field introduced by IDTs as laterally polarized with a uniform magnitude across the thickness of the plate. Thus, the integration of stress and the lateral electrical fields, which describes the mutual energy between the electrical and mechanical domains [12], vanishes for even-order modes and leads to zero electromechanical coupling to these modes (according to Berlincourt's definition of $k_t^2$). On the other hand, odd order modes can be excited due to non-zero integral of mutual energy. For the purpose of frequency scaling to Ka band, our overmoding approach will focus on the odd order modes with the simple top-only IDTs.

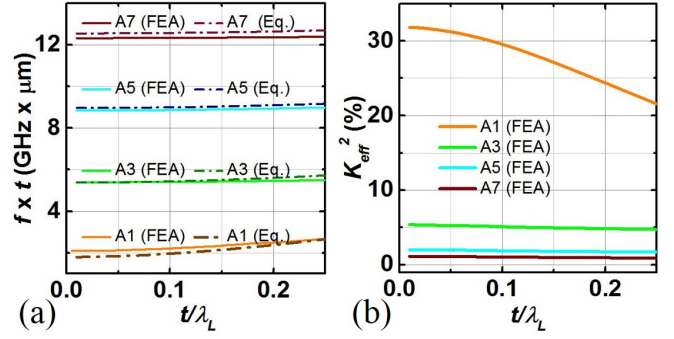

*Figure 3: (a) Calculated and simulated f×t product and (b) Simulated effective electromechanical coupling factor of A1, A3, A5, and A7 modes vs. the ratios of film thickness to longitudinal wavelength (h/λ$_L$).*

To more precisely predict the resonant frequency and electromechanical coupling of a higher odd order A-mode, we have to resort a more refined model that treats the cross-section of the resonator as a two-dimensional cavity, instead of an infinitely long slab seen in Fig. 1 [11]. The resonant frequency of an odd order mode in a two-dimensional cavity with thickness of $t$ and length of $l$ is given by

$$f_0^{mn} = \sqrt{(\frac{mv_t}{2t})^2 + (\frac{nv_L}{2l})^2} \quad (1)$$

where $m$ and $n$ are the mode orders in the vertical (z-axis) and longitudinal (x-axis) directions, respectively. $v_t$ and $v_L$ are the acoustic phase velocities in the vertical and longitudinal directions. The asymmetric modes of interest in this work have a longitudinal mode order $n$ of 1 with a vertical mode order m that takes a value among 1, 3, 5 and 7. Note that Eq. 1 assumes the lateral boundaries are mechanically free, the same as the top and bottom surfaces. This assumption will be revisited later. Eq. 1 also implies that any composite mode of order $f_0^{mn}$ with $n$ taking an odd value larger than 1 can emerge as a spurious mode near the intended mode $f_0^{m1}$. Such a phenomenon has been observed for other high coupling resonators with a 2D nature [13] [14].

For modes with $n=1$, $l$ is equal to half of the longitudinal wavelength ($\lambda_L/2$). Thus, Eq. 1 can be re-written as:

$$f_0^{m1} = \frac{v_L}{2t}\sqrt{(\alpha m)^2 + (2\frac{t}{\lambda_L})^2} \quad (2)$$

where $\alpha$ is the ratio between the velocity of vertical and longitudinal directions:

$$\alpha = \sqrt{c_{55}/c_{11}} \quad (3)$$

To validate the two-dimensional analysis, Comsol-based FEA is used to calculate the frequency variation of each odd-order asymmetrical mode in a Z-cut LiNbO$_3$ thin film. Fig. 3 (a) shows the simulated and calculated frequency-thickness products ($f \times t$) as a function of the ratio of plate thickness to longitudinal wavelength ($t/\lambda_L$). The discrepancies between simulation and calculation are likely caused by several simplifications used in the calculation.

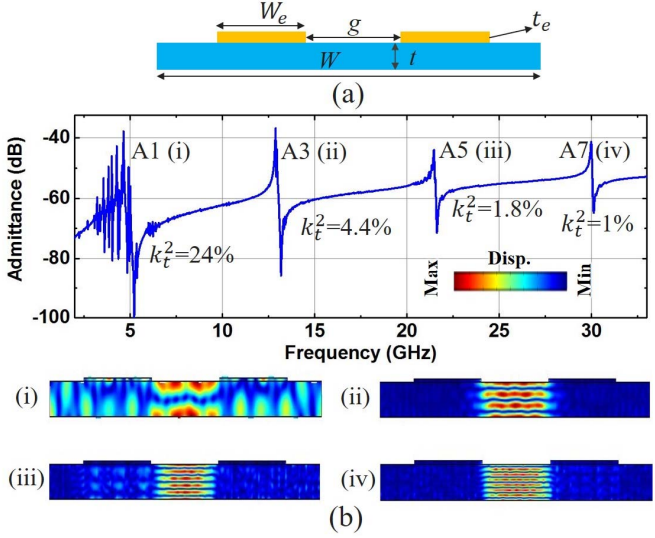

*Figure 4: (a) Mocked-up view of designed asymmetric mode LiNbO$_3$ resonator. (b) Simulated response of a Z-cut LiNbO$_3$ A-mode resonator. Displacement mode shapes of (i)A1, (ii)A3, (iii)A5, and (iv)A7 are included.*

TABLE I
Physical dimensions of the designed asymmetric mode resonator

| Electrode width, $W_e$ | Device width, $W$ | Electrode thickness, $t_e$ | Film thickness, $t$ | Spacing between electrodes, $g$ | Total length, $L$ |
|---|---|---|---|---|---|
| 3 μm | 13 μm | 60 nm | 400 nm | 3 μm | 100 μm |

In addition to the resonant frequency, each mode is also characterized by an effective electromechanical coupling factor ($k_{eff}^2$) [15]:

$$k_{eff}^2 \approx \frac{e^2}{\varepsilon^S c^E} \cdot k_{v\_m}^2 \cdot k_{l\_n}^2 \quad (4)$$

where $e$ is the piezoelectric efficient, $\varepsilon^S$ is the permittivity under constant strain, $c^E$ is the stiffness under constant electric field. $k_{v\_m}^2$ is a scaling factor capturing the dependence of $k_{eff}^2$ on the stress and electric field distributions of the $m^{th}$ order mode in the vertical direction. Similarly, $k_{l\_n}^2$ represents the dependence of $k_{eff}^2$ on the stress field distribution of $n^{th}$ order mode in the longitudinal direction [11]. The expression of $k_{v\_m}^2$, as part of the Berlincourt Formula, is given as:

$$k_{v\_m}^2 = \frac{(\int E_X(z) u_X(z) dz)^2}{\int E_X^2(z) dz \cdot \int u_X^2(z) dz} \quad (5)$$

By using the simplified field distributions shown in Fig. 2 for integration, $k_{v\_m}^2$ can be formulated as a function of mode order m:

$$k_{v\_m}^2 = \frac{1}{m^2} \quad (6)$$

$k_{l\_n}^2$ in Eq. 4 is dependent to the ratio between the vertical and longitudinal dimensions ($t/\lambda_L$), and typically increases with respect to $m$. Therefore, the lower bound for $k_{eff}^2$ of the $m^{th}$-order ($m >1$) asymmetric mode is $1/m^2$ of $k_{eff}^2$ of the A1 mode.

To further understand the diminishing effect of overmoding on $k_{eff}^2$ (also the dispersive relationship between $k_{l\_n}^2$ and $t/\lambda_L$), Comsol FEA is used to calculate the $k_{eff}^2$ as a function of $t/\lambda_L$.

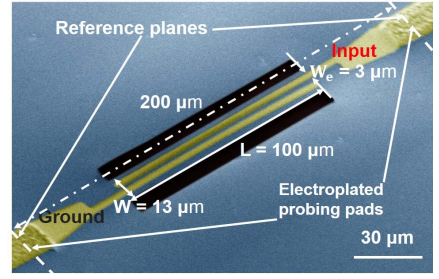

*Figure 5: SEM image of the fabricated device.*

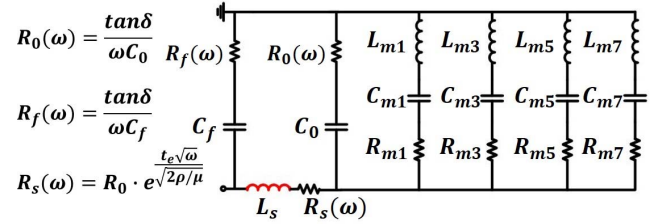

*Figure 6: Multi-resonance equivalent MBVD circuit model with an added series inductor. tanδ is the loss tangent of LiNbO$_3$, ρ is the resistivity of gold, and μ is the permeability of Au.*

The relationship, shown in Fig. 3(b), indicates that overmoding into A3, A5, and A7 can still have meaningful $k_{eff}^2$ for 5G applications if it starts with a large $k_{eff}^2$ of 30%, which is attainable for A1 mode with a $t/\lambda_L$<0.1.

Based on the above analyses (summarized in Fig. 3), we choose the thickness of the thin film to be 400 nm, and $\lambda_L$ to be 6 μm so that the targeted A5 and A7 modes can be scaled toward Ka band with reasonably large $k_{eff}^2$. $t$ of 400 nm and $\lambda_L$ of 6 μm are also deemed as a good tradeoff between having an adequately low $t/\lambda_L$ and achieving a reasonably large static capacitance ($C_0$) for the resonator. Based on Fig. 3 (a), the A7 mode in such a cavity would yield a resonant frequency at 30 GHz.

### III. DESIGN AND MODELING OF ASYMMETRIC MODE RESONATOR

To validate resonant characteristics and factor in the effects of electrodes, the resonator is simulated with 2D Comsol FEA. As shown in the cross-sectional mock-up in Fig. 4 (a), the device consists of a 2-electrode transducer on top of a mechanically suspended 400 nm thick Z-cut LiNbO$_3$ thin film. The two electrodes, connected to signal and ground respectively, induce lateral electric fields in the LiNbO$_3$ thin film, which subsequently excite the resonator into odd-order asymmetric mode vibration. 60 nm gold is used as the electrodes in the simulation. Gold is used for its good conductivity while thickness is kept low to avoid excess mass loading and shifting down the resonances. The design parameters are summarized in Table I. The Comsol simulated response is shown in Fig. 4 (b), including the displacement mode shape of each excited asymmetric mode. Different from fundamental Lamb waves (S0, A0), the higher order A-modes are confined between two electrodes, so the longitudinal cavity dimension is approximately the distance between adjacent electrodes. We already considered this effect in the design process by setting the distance between the electrodes to 3 μm, half of the target $\lambda_L$ value. This local confinement of displacement for A3, A5 and A7 modes is believed to be caused by the acoustic impedance mismatch between the metalized and unmetalized sections, which increases with respect to the mode order. Strictly speaking, Eq. 2 can no longer apply to calculating the resonances as such confinement does not have mechanical free boundaries assumed in the derivation of Eq. 2. Instead, complex reflection coefficients should be used to capture longitudinal boundaries. However, the tight confinement as seen in Fig. 4 leads to the belief that Eq. 2 could be still used for estimates of resonances with good accuracy.

TABLE II
Key measured values (in bold symbols) and extracted parameters of the multi-resonance MBVD model

| Order ($m$) | $f_0$ | $R_{mi}$ | $C_{mi}$ | $L_{mi}$ | $R_s(\omega)$ | $C_0$ | $C_f$ | $L_s$ | $\tan\delta$ | $Q_o$ | $k_t^2$ | FoM | Extracted Mechanical $Q_m$ |
|---|---|---|---|---|---|---|---|---|---|---|---|---|---|
| 1 | 4.5 GHz | 74 Ω | 3.84 fF | 319 nH | 12.5 Ω | | | | | 118 | 24% | 28.3 | 122 |
| 3 | 12.9 GHz | 78 Ω | 0.574 fF | 267 nH | 12.9 Ω | 19.4 fF | 13.5 fF | 90 pH | 0.07 | 224 | 3.7 % | 8.3 | 277 |
| 5 | 21.4 GHz | 82 Ω | 0.225 fF | 247.3 nH | 13.2 Ω | | | | | 287 | 1.5 % | 4.3 | 406 |
| 7 | 29.9 GHz | 76 Ω | 0.15 fF | 192 nH | 13.4 Ω | | | | | 328 | 0.94% | 3.1 | 474 |

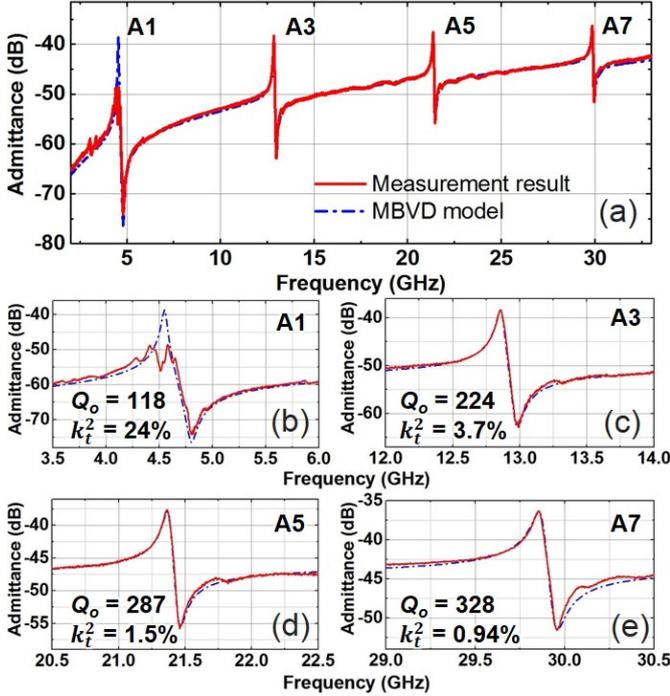

*Figure 7: (a) Measured and multi-resonant equivalent MBVD circuit modeled response of an asymmetric mode LiNbO₃ resonator with zoomed-in view of (b) A1, (c) A3, (d) A5, and (e) A7 mode resonances.*

Consistent with our theoretical analyses in Section II, FEA results show that only the odd-order modes are excited. The first four odd modes, A1, A3, A5 and A7, have resonant frequencies at 4.5, 12.8, 22, and 30 GHz respectively. The simulated effective electromechanical coupling factors for these modes are 24%, 4.4%, 1.8% and 1%, showing the theoretically predicted diminishing trend. It scales slightly better than $1/m^2$ due to the aforementioned dispersive relationship between $k_{L\_n}^2$ and mode order, the simulated results on resonator $k_t^2$ are also on par with the predictions based on Eq. 4 and in Fig. 3(b).

## IV. MEASUREMENTS AND DISCUSSIONS

To validate the analytical and modeling results, the designed Z-cut LiNbO₃ A-mode resonator was fabricated on a 400 nm-transferred Z-cut LiNbO₃ thin film following the process described in [9]. The fabricated device is shown in Fig. 5. 60 nm-Au is sputtered and lift-off as the top electrodes. The probing pads, also made of Au, are electroplated to 2.5 μm to reduce loss. To reduce the parasitic effect between the lead lines, a 200 μm pitch between probing pads is used. The fabricated device was characterized over a wide frequency range (up to 40 GHz) with a Keysight N5230A PNA-L network analyzer in dry air and at room temperature. A TRL calibration based on on-wafer standards is used to move the measurement reference planes to the positions as labeled in Fig. 5.

Similar to the simulation results in Fig. 4, the measurement results shown in Fig.7(a) exhibit resonant frequencies at 4.5, 12.8, 21.4, and 29.9 GHz, corresponding to the anticipated A1, A3, A5, and A7 modes respectively. As expected, even-order asymmetric modes cannot be observed with this design. Among the first four odd-order asymmetric modes shown in Fig. 7(b-d), a higher order yields a higher measured overall quality factor, $Q_o$ (including the effects of all loss mechanisms). The reason behind this phenomenon is still under investigation.

A multi-resonance modified Butterworth-Van Dyke (MBVD) model, in which each resonance is captured by a motional branch of $R_m$, $L_m$, and $C_m$, is used to interpret the measurement results. As shown in Fig. 6, an additional series inductor ($L_s$) is added to account for the non-negligible inductance of the electrode fingers at high frequencies. An additional resistor ($R_s$) is added to account for surface resistance of electrodes and bus lines, the value of which is dependent on frequency due to skin effects and calculated based on the equation shown in Fig. 6. To account for the parasitic effects, $C_f$ and $R_f$ are included as the feedthrough capacitance and loss in the substrate, respectively. An on-chip test structure consisting of only the bus lines was included in fabrication to measure the value of $C_f$. Except for the overall $Q_o$, resonances, and $C_f$, other parameters in the multi-resonance MBVD model are extracted from measurements. By excluding the loss in the electrical domain, we can also extract the mechanical $Q_m$ at these resonances from the MBVD model. It is worth noting that the A7 mode demonstrates a very high mechanical $Q$ of 474 at 30 GHz. $Q_m$ for other resonances, along with extracted $k_t^2$ of each mode and other key parameters of the MBVD models, are listed in Table II. As seen in Fig. 7, the response of the MBVD model excellently fits with measured admittance response, hence validating the parameter extraction. The effects of self-inductance and feedthrough capacitance have been de-embedded for the extracted $k_t^2$ values. The extracted $k_t^2$ values match the simulated results shown in Fig. 4, as well as the analytical results in Fig. 3(b). The agreement thus confirms the effectiveness of our frequency scaling approach based on overmoding asymmetric modes.

The measured results herein have demonstrated the potential of higher-order A-modes for enabling Ka band acoustic resonators. A mechanical $Q$ of 474, as demonstrated for A7, has already exceeded the state of the FBAR results [3]. Further $Q$ enhancement has to reply on more sophisticated understanding the acoustic loss at this frequency range, as well as reducing dissipation in the electrical domain. To achieve a large $k_t^2$ and higher filter bandwidth at these frequencies, a thinner LiNbO$_3$ film may be adopted with our approach to scale the A5 mode to Ka band. This, however, will likely comes at the cost of lower power handling and linearity.

## V. CONCLUSION

In this work, extending asymmetric modes in a LiNbO$_3$ thin film to higher orders is first theoretically studied for the purpose of scaling asymmetric mode resonances toward Ka band. Subsequently, a LiNbO$_3$ MEMS resonator with various resonances, the highest at 29.9 GHz, has been demonstrated for the first time. The developed MEMS resonator exhibits a $k_t^2$ of 0.94%, a measured overall $Q_o$ of 328, and an extracted mechanical $Q$ of 474 at 29.9 GHz. Further development and optimization of LiNbO$_3$ asymmetric resonators could lead to miniature Ka-band acoustic front-end filter technology that is currently unavailable for 5G.